\documentclass[aps, prl, amsmath, amssymb, superscriptaddress, reprint]{revtex4-1}

\usepackage{graphicx}

\begin{document}

\title{Optimization of sharp and viewing-angle-independent structural color}

\author{Chia Wei Hsu}\email{cwhsu@mit.edu}
\affiliation{Department of Physics, Massachusetts Institute of Technology, Cambridge, Massachusetts 02139, USA}
\affiliation{Department of Physics, Harvard University, Cambridge, Massachusetts 02138, USA}

\author{Owen D. Miller}
\affiliation{Department of Mathematics, Massachusetts Institute of Technology, Cambridge, Massachusetts 02139, USA}

\author{Steven G. Johnson}
\affiliation{Department of Mathematics, Massachusetts Institute of Technology, Cambridge, Massachusetts 02139, USA}

\author{Marin Solja\v{c}i\'{c}}
\affiliation{Department of Physics, Massachusetts Institute of Technology, Cambridge, Massachusetts 02139, USA}

\begin{abstract}
Structural coloration produces some of the most brilliant colors in nature and has many applications.
However, the two competing properties of narrow bandwidth and broad viewing angle have not been achieved simultaneously in previous studies.
Here, we use numerical optimization to discover geometries where a sharp 7\% bandwidth in scattering is achieved, yet the peak wavelength varies less than 1\%, and the peak height and peak width vary less than 6\% over broad viewing angles (0--$90^\circ$) under a directional illumination.
Our model system consists of dipole scatterers arranged into several rings; interference among the scattered waves is optimized to yield the wavelength-selective and angle-insensitive response. Such designs can be useful for the recently proposed transparent displays that are based on wavelength-selective scattering.
\end{abstract}

\maketitle

%%%%%%%%%%%%%%%%%%%%%%%%%%  body  %%%%%%%%%%%%%%%%%%%%%%%%%%
\section{Introduction}

Wavelength-scale structures can modify the spectrum of light scattering through interference, giving rise to coloration in many natural structures~\cite{2008_Kinoshita_RPP, 1999_Vukusic_PRSB,1998_Prum_Nature, 2009_Whitney_Science} and leading to a variety of applications~\cite{2007_Ozin_nphoton, 2007_Potyrailo_NP, 2009_Kim_nphoton}.
The recently proposed transparent displays based on narrow-band light scattering~\cite{2014_Hsu_ncomms} represent an unexplored opportunity for such interference-based structural color: 
structures with a wavelength-selective and angle-insensitive scattering response can be embedded in a transparent medium to create a screen that is transparent to the ambient light but capable of displaying images projected from a narrow-band light source.
For transparency and for wide viewing angle, one must simultaneously achieve the competing properties of wavelength selectivity and viewing-angle insensitivity, under directional rather than omnidirectional illumination.
Here, we show with numerical calculation that a narrow 7\% bandwidth can be obtained for scattering at a broad range of observation angles (0--$90^\circ$) with directional illumination of an optimized structure---this is a significant improvement over the previous design using plasmonic particles~\cite{2014_Hsu_ncomms}, where the bandwidth was about 20\% and there was undesirable absorption that reduced transparency.
In addition to the bandwidth, the peak wavelength and peak intensity of scattering are also insensitive to the viewing angle.
The structure consists of a collection of wavelength-scale ring scatterers amenable to fabrication by direct laser writing (multiphoton lithography)~\cite{2004_Deubel_nmat, 2007_Haske_OE, 2009_Li_Science, 2010_Szameit_JPB, 2014_Bauer_PNAS} and is optimized so that constructive interference occurs only in a narrow bandwidth but with a dipole-like broad-angle pattern.
We optimize the structure as modeled by semi-analytical scattering theory, and validate the results via 3D boundary-element method (BEM) simulations.

One route to narrow-band broad-angle scattering is through resonances in plasmonic particles~\cite{2014_Hsu_ncomms, 2014_Hsu_NL}, but that approach is limited by the absorption loss of metal: for the response to be dominated by scattering rather than absorption, the particle must have a radiation loss higher than the absorption loss~\cite{2007_Hamam_PRA, 2010_Ruan_JPCC, 2014_Miller_PRL}, making the total loss high and the resonance broad.
Dielectric particles have no absorption, but subwavelength dielectric particles lack wavelength selectivity, and larger dielectric particles have multiple resonances overlapping each other~\cite{2010_SiNW_color_NL, 2014_Hsu_NL}.
In this paper, we explore an alternative route based on interference effects in structural coloration, which is not constrained by the above-mentioned limitations. %but achieving wavelength selectivity and angle independence simultaneously is challenging.
Previous work on synthetic structural color has explored a wide range of designs, but none achieves narrow bandwidth and broad viewing angle simultaneously under directional illumination.
Multilayer films~\cite{2009_Bonifacio_AM, 2013_Kats_NM, 2014_Shrestha_srep}, periodically-modulated surfaces~\cite{2009_Cheong_APL, 2013_Yang_OE, Munk_book, 2014_Shen_arXiv}, and three-dimensional photonic crystals~\cite{JJ_book} reflect light only at a discrete set of angles rather than omnidirectionally.
Structural randomness can decrease the angular dependence and is generally understood to be the reason for the non-iridescence of Morpho butterfly's blue color~\cite{2009_Lee_AO, 2011_Saito_JNN, 2012_Steindorfer_OE, 2012_Chung_AM}, but the randomization also broadens the bandwidth to about 100 nm or larger.
Similarly, amorphous structures have weak angular dependence~\cite{2009_Dufresne_SoftMat, 2009_Takeoka_AMI, 2009_Ueno_ChemComm, 2010_HarunUrRashid_CPC, 2010_Forster_AM, 2010_Noh_AM, 2012_Takeoka_JMC_review, 2012_Magkiriadou_OME, 2012_Saranathan_JRSI, 2014_Magkiriadou_arXiv}
but with spectra that are broad and with viewing-angle-dependent peak wavelengths when illuminated directionally~\cite{2010_Noh_AM}.
Wavelength selectivity and angle insensitivity are hard to achieve simultaneously because interference effects couple the dependence on wavelength and the dependence on angle.
Also, the majority of the prior studies focused on the physiologically perceived color of surfaces under broadband and diffuse omnidirectional illumination, where a narrow bandwidth is not necessary since the human-eye color matching functions are broad with widths of about 100 nm~\cite{color_book}, and where the diffuse illumination reduces the viewing-angle dependence at the expense of a broader bandwidth.
For example, topological optimization has been used to design perceived colors of surfaces~\cite{2014_Andkjaer_JOSAB, 2014_Johansen_JOSAB}, and the spectrum was not optimized to be narrower than 100 nm.
%A survey of the literature suggests that existing structural color designs are not appropriate for the new application on transparent display.
To explore the extent to which structural color can be useful for the new application on transparent display, here we carry out an optimization-based study.
%The new application on transparent display requires properties not achieved in the prior studies on structural color, and to explore the extent to which structural color can be useful for this application, here we carry out an optimization-based study.

\section{Light scattering and structure factor}

We consider light scattering from a collection of point scatterers where multiple scattering is negligible.
We start with the more general problem considering arbitrary configurations of points, and later in section~\ref{sec:ring}, motivated by the results of the arbitrary-structure optimization, we specialize to a smaller design space with point scatterers arranged into stacked rings.
When multiple scattering is negligible, the optical response is decoupled into a form factor that captures light scattering from an individual particle, and a structure factor that captures the effect of interference among waves scattered from different particles.
For small particles where the electric-dipole scattering dominates, the differential scattering cross section from this collection is~\cite{Jackson_book}
\begin{equation}
\label{eq:sigma}
\frac{d \sigma}{d \Omega} = k^4 \left| \hat{\bf e}_{\rm in} \cdot \hat{\bf e}_{\rm out}^* \right|^2 S({\bf q}),
\end{equation}
where $k = 2 \pi /\lambda$ is the wavenumber, $\lambda$ is the wavelength, $\hat{\bf e}_{\rm in}$, $\hat{\bf e}_{\rm out}$ are the polarization vectors of the incident and the scattered light, and the structure factor is given by
\begin{equation}
\label{eq:sq}
S({\bf q}) = \left| \sum_{j} \alpha_j e^{i {\bf q} \cdot {\bf r}_j} \right|^2.
\end{equation}
Here, $\alpha_j$ is the electric polarizability  of the $j$-th scatterer, ${\bf r}_j$ is its position, and ${\bf q}$ is the momentum transfer vector
\begin{equation}
\label{eq:q}
{\bf q} = k (\hat{\bf n}_{\rm in} - \hat{\bf n}_{\rm out}),
\end{equation}
with $\hat{\bf n}_{\rm in}$ and $\hat{\bf n}_{\rm out}$ being the propagation directions of the incident and the scattered light.
Note that we include $\alpha_j$ in the definition of $S({\bf q})$ to account for different types of scatterers; each scatterer is assumed to be a dielectric scatterer with a frequency-independent $\alpha_j$.
%, but we do not account for material dispersion ({\it i.e.}, no frequency dependence in $\alpha_j$).
In this model system, the angle and wavelength dependence comes primarily from $S({\bf q})$, which can be calculated efficiently for fast optimization.
In section~\ref{sec:BEM}, we validate
the assumptions in this model---that multiple scattering is negligible and that dipole scattering dominates with a constant scalar polarizabilty---with full-wave simulations.

The interference may provide the desired wavelength selectivity in $S({\bf q})$.
However, this wavelength 
selectivity typically comes with angle dependences that are undesirable, because wavelength and angles are coupled in the momentum transfer ${\bf q}$ [Eq.~(\ref{eq:q})].
For example, periodic structures scatter strongly when ${\bf q}$ lies on the reciprocal lattice~\cite{Jackson_book}, and the peak wavelength depends sensitively on the angles.
The $S({\bf q})$ of amorphous structures is a function of $|{\bf q}| = 2 k \sin(\theta/2)$, so the peak wavelength is a function of the angle $\theta$ between $\hat{\bf n}_{\rm in}$ and $\hat{\bf n}_{\rm out}$~\cite{2010_Noh_AM}; for a fixed illumination direction, the peak wavelength varies with the viewing angle.

\section{Optimization\label{sec:opt}}

\begin{figure}[tb]
   \centerline{
   \includegraphics[width=1.7in]{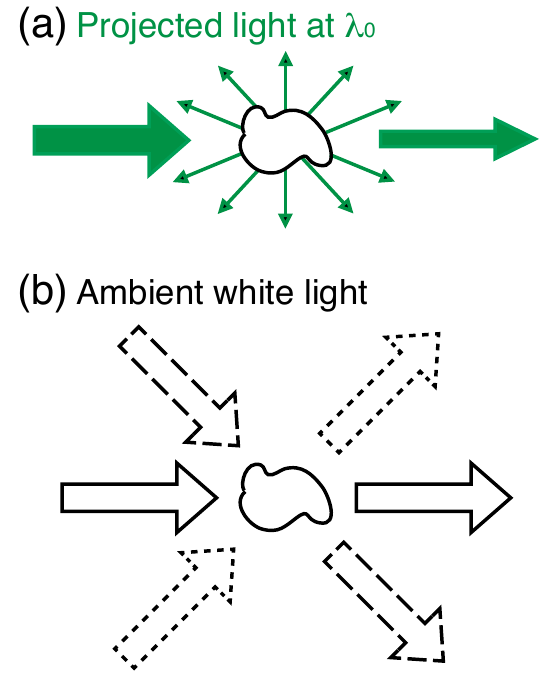} 
   }
   \caption{Illustration of the desired scattering characteristics. (a) Light coming from the projector, which is monochromatic at wavelength $\lambda_0$ and incident from a specific direction $\hat{\bf n}_{\rm in}$, is scattered strongly into all viewing angles. (b) Ambient white light, which is broad band and incident from all directions (illustrated with different arrows), passes through with little scattering.}  
   \label{fig:fig1}
\end{figure}

Figure 1 illustrates the desired scattering characteristics.
At the target wavelength $\lambda_0$, the structure scatters light strongly and uniformly into all potential viewing angles.
At other wavelengths, the structure scatters light weakly so it appears transparent to the ambient white light.
In practice, many copies of the same structure will be distributed randomly and evenly in a transparent medium (such as a polymer film) to create a ``screen'' that is mostly transparent to the ambient light but capable of displaying images projected onto it at the target wavelength $\lambda_0$~\cite{2014_Hsu_ncomms}.

To search for structures whose structure factor $S({\bf q})$ exhibits such scattering characteristics, we define a figure of merit (FOM),
\begin{equation}
\label{eq:fom}
{\rm FOM} = \frac{A}{B+C},
\end{equation}
where $A$ is $S({\bf q})$ at $\lambda_0$ averaged over the viewing angles $\hat{\bf n}_{\rm out}$,
$B$ is $S({\bf q})$ averaged over both the viewing angles $\hat{\bf n}_{\rm out}$ and wavelengths in the visible spectrum, and $C$ is the maximum value (within the visible spectrum) of the standard deviation of $S({\bf q})$ with respect to the viewing angles.
High FOM can only be achieved when $A$ is large (strong scattering at $\lambda_0$), $B$ is small (weak scattering at other wavelengths), and $C$ is small (independence of the viewing angle).
Similar to Ref.~\cite{2014_Hsu_ncomms}, we define the FOM as a ratio; the absolute value of scattering strength per structure is of less interest because stronger response can be achieved with a higher concentration of the structures.
Here, we consider the scenario where images are projected from the $z$ direction and viewers view from various directions on the same side.
So, the calculation of $A$ and $C$ takes $\hat{\bf n}_{\rm in} = \hat{\bf z}$ as fixed and integrates
over all solid angles $\hat{\bf n}_{\rm out}$ with $\hat{\bf n}_{\rm out} \cdot \hat{\bf z} \le 0$.
For high transparency at all angles, the calculation of $B$ should integrate over all possible $\hat{\bf n}_{\rm in}$, all possible $\hat{\bf n}_{\rm out}$, and all wavelengths in the spectrum of interest; however, since $S({\bf q})$ is only a function of ${\bf q} = k (\hat{\bf n}_{\rm in} - \hat{\bf n}_{\rm out})$, one of the averages is redundant, and in our implementation we calculate $B$ by integrating over $\hat{\bf n}_{\rm out}$ and the wavelength only.

Due to the high dimensionality of the parameter space and the unevenness of the FOM landscape, directly applying global or local optimization algorithms tends to yield results that are very sensitive on the initial guess and typically very suboptimal.
Therefore, we employ a step-wise search strategy~\cite{2007_Lu_JOSAB, Miller_thesis}:
starting from vacuum, we add components one by one. At each step, a global optimization is performed to find parameters of the new component, followed by a local optimization on the parameters of all components.
This procedure does not enumerate all possible configurations (which would be impossible computationally), but it keeps the FOM high, effectively providing a good initial guess in the high-dimensional parameter space.
% as the parameter space expands its dimensionality.
After some experimentation with a free optimization package~\cite{NLOPT}, we chose well known local~\cite{2002_Svanberg_MMA} and global~\cite{2006_Kaelo_CRS} search algorithms and implemented our FOM [Eq.~(\ref{eq:fom})] to four-digit accuracy in the high-performance dynamic language Julia~\cite{Julia}.
The best structure is picked from results of several independent runs.

\section{Ring-based structures\label{sec:ring}}

We compare optimization results in the broader search space where all possible configurations are allowed, and the smaller search space where the structures consist of several rings aligned along the $z$ axis. We find that with sufficient number of points ($N \gtrsim 100$), the narrower search leads to FOM comparable to or sometimes higher than the highest FOM found in the broader search.
This is because the ring-based structures eliminate the azimuthal part of the viewing-angle dependence and have lower $C$ in general.
In addition, the reduction of the search space makes the optimization more efficient and avoids the poor local optima in the broader search space.
With this insight, we focus our attention on structures made of rings aligned along the $z$ axis.
Instead of parameterizing coordinates of the individual point scatterers, we now directly parameterize coordinates of the rings.
The computation is further accelerated by analytically summing over point scatterers in a ring (assuming closely spaced points), as
\begin{equation}
\label{eq:sq_ring}
S({\bf q}) = \left| \sum_{j} w_j e^{i q_z z_j} J_0(q_\rho \rho_j) \right|^2,
\end{equation}
where the new summation is over the constituent rings, $q_z$ and $q_\rho$ are the $z$ and the radial components of the ${\bf q}$ vector, and $J_0$ is the Bessel function. Each ring is specified by three parameters: its $z$-coordinate $z_j$, its radius $\rho_j$, and its weight $w_j$ (which is given by the sum of polarizability among the point scatterers in this ring).
A ring with a very small radius $\rho_j$ is effectively a point scatterer on the $z$ axis.

\begin{figure}[t]
   \centerline{
   \includegraphics[width=3.5in]{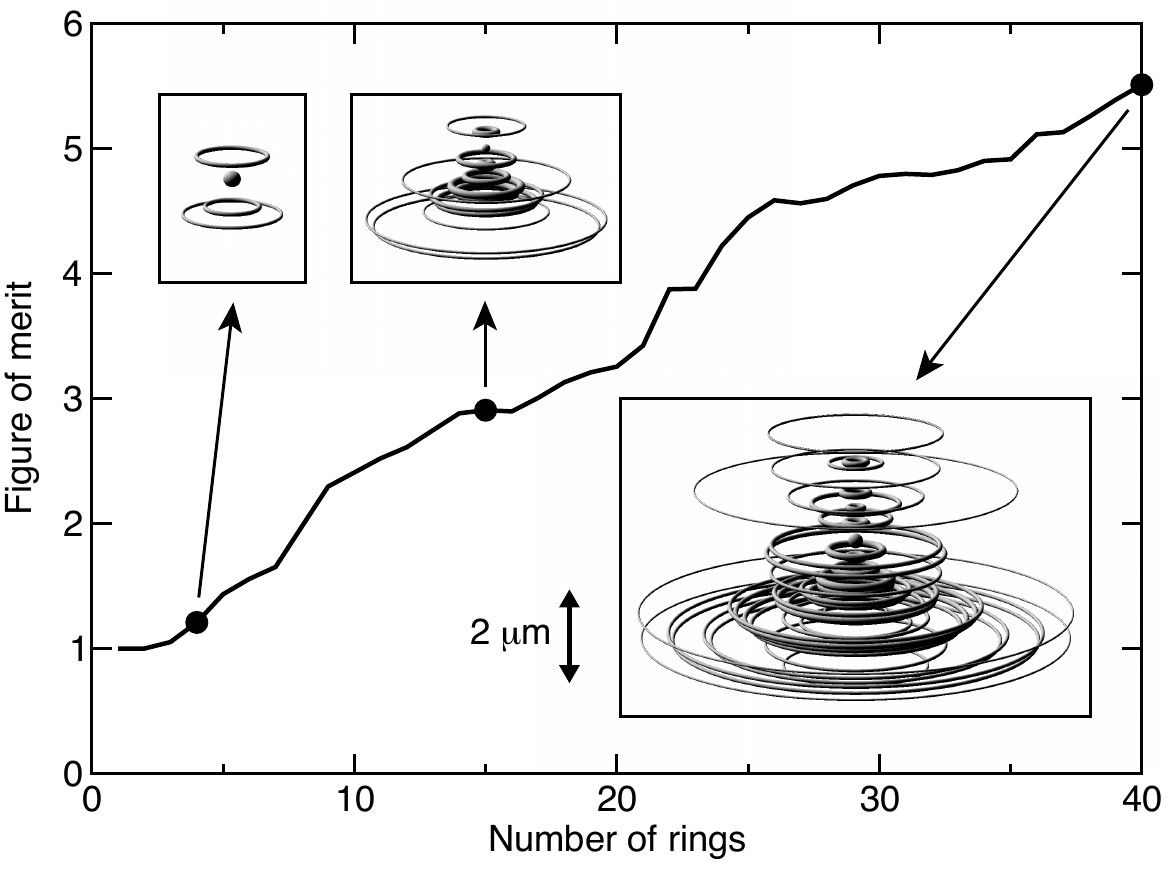} 
   }
   \caption{
   The evolution of the FOM for one instance of the step-wise optimization procedure. Starting from vacuum, rings are added one by one, with global optimization used to find parameters of the new ring and local optimization used to relax the whole structure.
   Insets show the intermediate and final configurations with 4, 15, and 40 rings, all drawn to the same length scale as indicated. Each ring is visualized as a torus with major radius $\rho_j$ and volume proportional to weight $w_j$, positioned at height $z_j$; all insets are drawn to the same length scale.
   }
   \label{fig:evoluation}
\end{figure}

\section{Results}

For the results presented here, we choose the wavelength window to be 400--800 nm and the target wavelength $\lambda_0$ to be 600 nm; structures for other target wavelengths (with corresponding shift in the window) can be obtained through scaling the structure sizes since Eq.~(\ref{eq:sigma}) is scale invariant.

\begin{figure*}[t]
   \centerline{
   \includegraphics[width=5.2in]{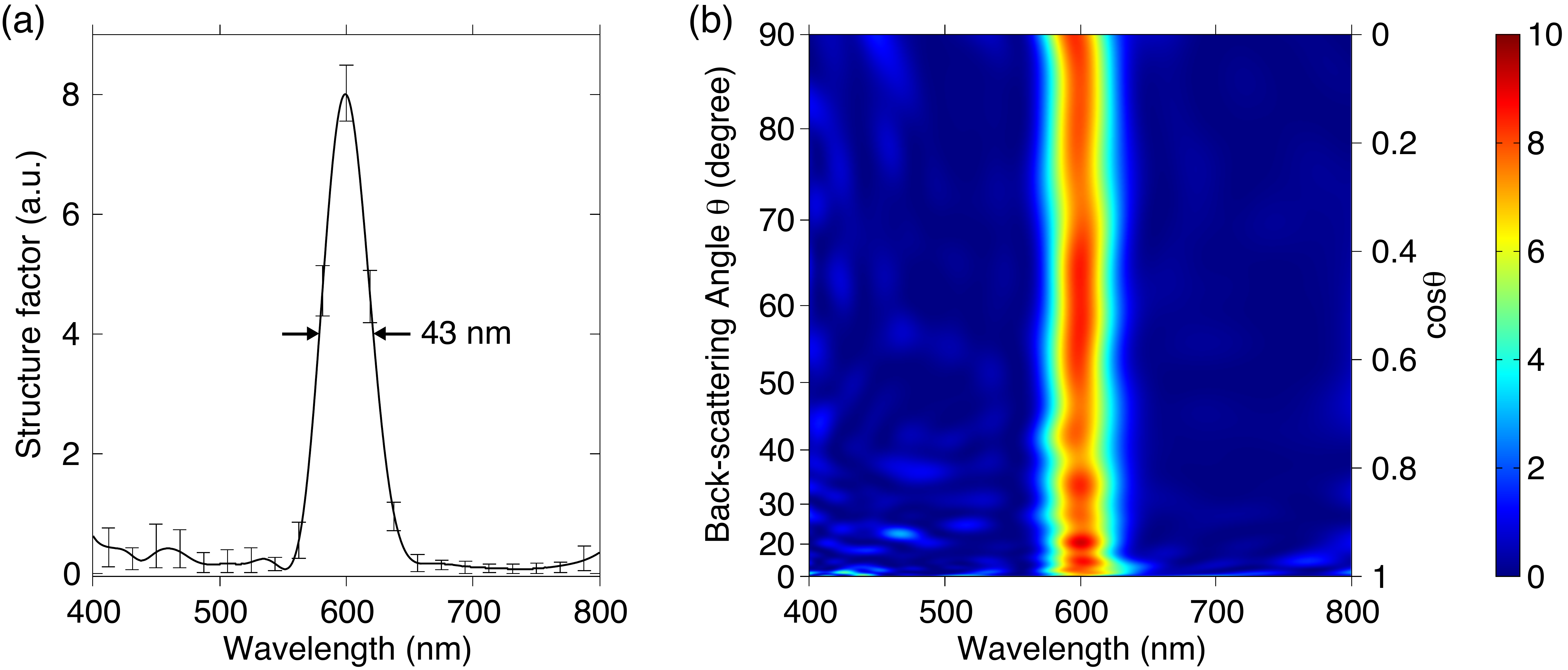} 
   }
   \caption{
   Structure factor $S({\bf q})$ for the 40-ring configuration optimized for wavelength-selective scattering at $\lambda_0 = 600$ nm and viewing-angle independence across 0 to 90 degrees.
   (a) Average value (solid line) and the 10th and the 90th percentiles (bars) with respect to the viewing angles.
   (b) Scattering-angle-resolved spectrum of $S({\bf q})$, for light incident along the $z$-axis. The $y$-axis scale accounts for the weight in the solid angle integration, $|\sin\theta {\rm d} \theta| = | {\rm d}(\cos\theta) |$.
   }
   \label{fig:result_sq}
\end{figure*}

Figure~\ref{fig:evoluation} shows the evolution of the FOM for one instance of the step-wise optimization procedure. As rings are added, the FOM grows steadily from 1 to higher than 5. The insets of Figure~\ref{fig:evoluation} show a few configurations during this process. In Supplementary Movie S1, we provide an animation showing the evolution of the configurations, of the scattering-angle-resolved spectrum of $S({\bf q})$, and of the FOM as rings are added; one can see the evolution of $S({\bf q})$ as it develops the wavelength selectivity while keeping the angular dependence low.

In Figure~\ref{fig:result_sq}, we plot the structure factor $S({\bf q})$ of the optimized configuration with 40 rings (the final structure shown in Figure~\ref{fig:evoluation} and Movie S1); Figure~\ref{fig:result_sq}(a) shows the angular mean and angular variation of $S({\bf q})$ as a function of wavelength, and Figure~\ref{fig:result_sq}(b) plots $S({\bf q})$ as a function of both the wavelength and the viewing angle.
The wavelength selectivity and angle independence are evident.
This $S({\bf q})$ reaches FOM = 5.51.
Its wavelength selectivity is characterized by $A/B = 7.64$; the full width at half maximum (FWHM) of its peak at $\lambda_0$ is only 43 nm, corresponding to a narrow 7\% bandwidth.
Its viewing-angle independence is characterized by $A/C = 19.8$; in the range of viewing angles considered (0--$90^\circ$), its peak wavelength has an angular mean and standard deviation of 599 $\pm$ 1.3 nm (0.2\% variation), its FWHM 43 $\pm$ 2.5 nm (6\% variation), and its peak value 8.0 $\pm$ 0.4 (5\% variation)---all highly insensitive to the viewing angle.
This scattering response is significantly better than the previous design using dilute concentration of spherical plasmonic particles~\cite{2014_Hsu_ncomms}, where the angular insensitivity was given by the spherical symmetry, yet absorption of the metal reduced transparency and limited the bandwidth of the scattering cross section to be 19\%, 22\%, and 24\% respectively for scattering of blue, green, and red lights.

The optimized parameters are tabulated in supplementary Tables S1. We see that the optimized structures do not exhibit regular patterns as grating-based structures do.
Given that our search procedure does not enumerate all possible configurations, we cannot rule out the possibility that there exist structures with regular patterns that perform equally well or better than the structures we found.
However, we find that imposing preference for regularity (such as linearly increasing $z$ coordinates or radii) during the optimization generally leads to suboptimal structures with much stronger angle dependence. This suggests that the irregularities are necessary for the low angle dependence, similar to the role of randomness in height- or position-randomized surfaces~\cite{2009_Lee_AO, 2011_Saito_JNN, 2012_Steindorfer_OE, 2012_Chung_AM}, although here the irregularities are not random but are optimized to produce the sharp peak in wavelength.
%We also note that the optimization procedure discovers structures with very different parameters but comparable performance, suggesting that we may be able to pick structures more amendable to fabrication without sacrificing much performance.

\section{Full-wave simulation of dielectric rings\label{sec:BEM}}

Our model system can be realized as dielectric rings embedded in a transparent medium, with low refractive-index contrast between the two.
Such a structure may be fabricated using direct laser writing (multiphoton lithography), which has been used to fabricate waveguides~\cite{2010_Szameit_JPB}, photonic crystals~\cite{2004_Deubel_nmat}, and many complex high-resolution three-dimensional structures~\cite{2014_Bauer_PNAS} with feature sizes as small as 40 nm~\cite{2007_Haske_OE, 2009_Li_Science}.
To cover a large-area surface, one may place many copies of the optimized structure at random positions on the surface to increase the overall response while avoiding inter-structure interference: when there is no correlation between the positions of the individual structures, the inter-structure cross terms in $S({\bf q})$ average to zero, and the total response is the per-structure response times the number of structures.

\begin{figure*}[t]
   \centerline{
   \includegraphics[width = 6 in]{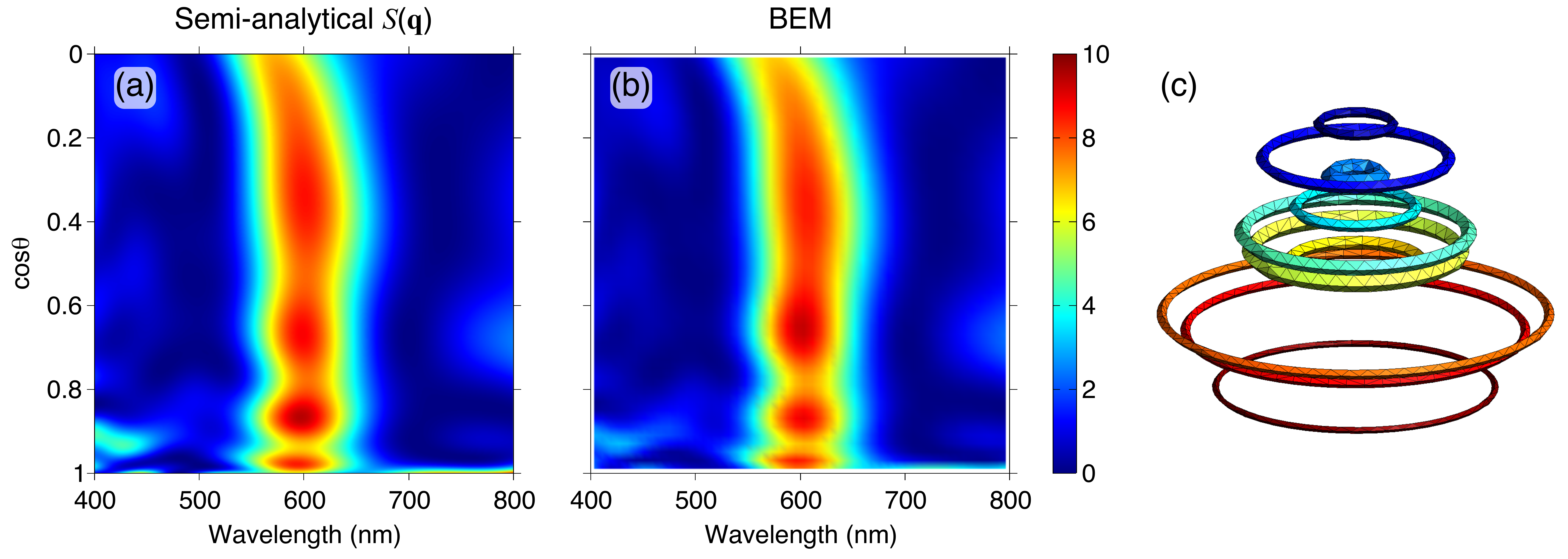} 
   }
   \caption{Comparison between the $S({\bf q})$ model and the exact scattering response calculated using the boundary-element method (BEM) for a 10-ring configuration.
   (a) Scattering-angle-resolved spectrum of $S({\bf q})$ (in arbitrary units).
   (b) Normalized differential scattering cross section, $(d\sigma/d\Omega)/k^4$ (in arbitrary units), calculated using BEM for the corresponding system of dielectric rings ($\epsilon=1.2$) in air with incident light polarized perpendicular to the scattering plane.
   (c) The surface mesh used in the BEM calculation.
   }
   \label{fig:BEM}
\end{figure*}

To verify that all assumptions in our model are valid in such a continuous dielctric-ring structure, we use a free-software implementation~\cite{2013_Reid_arxiv, scuff_EM} of the boundary-element method (BEM) to solve the corresponding scattering problem in the original 3D vectorial Maxwell's equations, with the differential scattering cross section calculated using Eq.~(10.93) of Ref.~\cite{Jackson_book} implemented within the BEM framework.
BEM employs no approximation aside from surface discretization, so it accounts for effects not considered in our model such as multiple scattering and scattering beyond the dipole approximation. 
Given the computational cost, we only perform the full-wave BEM calculation on a 10-ring structure, at 50 wavelengths and 50 angles.
We mesh the surfaces of the 10 rings with 3100 triangles [shown in Figure~\ref{fig:BEM}(c)], which provides a reasonable balance between accuracy and computation time.
%We take the incident light to be ${\bf E}_0 = E_0 e^{i k z} \hat{\bf x}$ and calculate light scattered into the $y$-$z$ plane (perpendicular to the incident polarization)
Here, we consider dielectric rings with $\epsilon=1.2$ in air.
The $z$ coordinate and major radius of each ring are taken directly from the already-optimized parameters $z_j$ and $\rho_j$. The weight $w_j$ is the total polarizability in each ring, so it is proportional to the ring volume; therefore, we choose the thickness of each ring to have its volume proportional to the optimized $w_j$.
%The actual index contrast used in an experiment will depend on the material and the fabrication procedure; higher index contrast provides larger polarizability and stronger scattering per structure, although its exact scattering response may differ more from the simplified $S({\bf q})$ model.
The thickness (minor diameter) of the rings lies between 42 nm and 122 nm. %, larger than the resolution limit of direct laser writing.

The angle-resolved scattering spectrum calculated using BEM is shown in Figure~\ref{fig:BEM}(b); it agrees very well with the prediction using structure factor $S({\bf q})$ [shown in Figure~\ref{fig:BEM}(a)], verifying that our model is appropriate for such low-index-contrast continuous-ring structures and that the optimized result has a certain degree of robustness with respect to how the model is realized.
The BEM result (without further optimization) corresponds to FOM = 2.51 (with $A/B = 3.58$ and $A/C = 8.38$), versus FOM = 2.54 (with $A/B = 3.51$ and $A/C = 9.15$) in the $S({\bf q})$ model where the optimization was performed.

\section{Conclusion}

We have provided an optimization approach based on a semi-analytical model of light scattering that reveals structures with narrow-band scattering that is independent of the viewing angle under directional illumination.
This wavelength-selective and viewing-angle-independent light scattering can be potentially useful for the recently proposed transparent display based on wavelength-selective scattering~\cite{2014_Hsu_ncomms}.
%Our design may also be useful for chemical and biosensing, as the optical response can be modified by the change of refractive indices.

The optimization-based design can be used for any given fixed illumination direction. However, we note that under directional illumination, simultaneous independence of both the incident and the outgoing angles is not possible in this model of scattering based on $S({\bf q})$; $S({\bf q})$ is a function of ${\bf q} = k (\hat{\bf n}_{\rm in} - \hat{\bf n}_{\rm out})$, so independence of both angles would require independence of the wavelength as well, meaning no wavelength selectivity. 
It may be possible, however, to reduce angle dependence or enhance wavelength selectivity by intentionally going beyond dipole scattering and single scattering, or by introducing resonances in the polarizabilities or in the form factors of individual scatterers; this may be the topic of future investigations.

A straightforward extension can consider structures that scatter light at multiple narrow bands, which can be useful for full-color transparent displays.
Future works can also explore structures with larger refractive index contrast, which can produce stronger scattering per structure and may require modeling beyond the $S({\bf q})$ analysis.
One may look for results even more insensitive to errors and implementation details via the robust-optimization techniques~\cite{2007_Bertsimas_JAP, 2009_Sigmund_AMS, 2009_Mutapcic_EO, 2011_Wang_JOSAB, 2012_Elesin_PN, 2012_Oskooi_OE, 2014_Men_OR, 2014_Men_OE}.
Another interesting future direction would be to explore whether there is a fundamental lower limit on the product of the angular and frequency bandwidths per unit volume, analogous to similar limits on the delay--bandwidth product~\cite{2007_Miller_PRL}.

\section*{Acknowledgments}

We thank Hui Cao, Seng Fatt Liew, and Yichen Shen for helpful discussions.
This work was supported in part by the U.~S.~Army Research Laboratory and the U.~S.~Army Research Office through the Institute for Soldier Nanotechnologies under contract number W911NF-13-D-0001, in part by the MIT Deshpande Center for Technological Innovation, and in part by the Materials Research Science and Engineering Centers of the National Science Foundation under grant No.~DMR-0819762, 

\bibliography{mybib}

%\clearpage

% Change the caption heading to Supplementary Figure X
\renewcommand{\theequation}{S.\arabic{equation}}
\renewcommand{\thefigure}{{S\arabic{figure}}}
\renewcommand{\thetable}{{S\arabic{table}}}

%\clearpage

\vspace{40pt}

\noindent {\bf \large Supplementary Tables}

\begin{table}[h]
\caption{Parameters of the optimized 40-ring structure shown in Figure 2 and Figure 3 of the main text.}
\begin{center}
\begin{tabular}{c @{\hspace{12pt}} c @{\hspace{12pt}} c @{\hspace{12pt}} | @{\hspace{12pt}} c @{\hspace{12pt}} c @{\hspace{12pt}} c}
$z_j$ ($\mu$m) & $\rho_j$ ($\mu$m) & $w_j$ (a.u.) & $z_j$ ($\mu$m) & $\rho_j$ ($\mu$m) & $w_j$ (a.u.) \\
\hline  
\hspace{0in} 2.380 & 1.714 & 0.375 &   -0.795 & 1.080 & 1.401 \\
\hspace{0in} 1.794 & 0.217 & 0.413 &   -0.868 & 1.621 & 1.803 \\
\hspace{0in} 1.776 & 0.526 & 0.352 &   -0.902 & 0.955 & 1.864 \\
\hspace{0in} 1.677 & 1.651 & 0.418 &   -0.999 & 1.566 & 2.706 \\
\hspace{0in} 1.257 & 3.187 & 0.506 &   -1.109 & 1.350 & 1.241 \\
\hspace{0in} 1.171 & 0.261 & 0.615 &   -1.137 & 2.515 & 2.056 \\
\hspace{0in} 1.103 & 1.335 & 0.533 &   -1.207 & 4.329 & 0.836 \\
\hspace{0in} 0.954 & 0.870 & 0.947 &   -1.271 & 0.621 & 0.279 \\
\hspace{0in} 0.872 & 0.154 & 0.238 &   -1.272 & 2.399 & 3.083 \\
\hspace{0in} 0.650 & 0.699 & 0.880 &   -1.394 & 1.448 & 1.003 \\
\hspace{0in} 0.549 & 0.205 & 0.685 &   -1.399 & 2.220 & 3.024 \\
\hspace{0in} 0.195 & 0.000 & 0.322 &   -1.488 & 1.099 & 0.742 \\
\hspace{0in} 0.100 & 1.678 & 1.678 &   -1.526 & 2.095 & 1.962 \\
\hspace{0in} 0.002 & 0.543 & 1.309 &   -1.647 & 3.741 & 2.208 \\
-0.056 & 1.774 & 1.880 &               -1.702 & 3.036 & 1.747 \\
-0.138 & 0.163 & 0.486 &               -1.727 & 4.281 & 0.834 \\
-0.377 & 0.426 & 0.939 &               -1.795 & 3.657 & 2.190 \\
-0.457 & 1.686 & 1.054 &               -1.835 & 2.899 & 1.649 \\
-0.578 & 0.714 & 2.231 &               -1.939 & 1.807 & 0.947 \\
-0.706 & 0.692 & 2.225 &               -2.380 & 1.448 & 0.541 \\
\hline
\end{tabular}
\end{center}
\end{table}

\end{document}